\documentclass[a4paper,11pt]{article}
\usepackage{jcap/jcap}
\usepackage{lineno}
\usepackage{amsmath}
\usepackage{graphicx}
\usepackage{subcaption}
\usepackage{gensymb}
\usepackage{xcolor}
\usepackage{orcidlink}

\newcommand{\revr}[1]{{\color{black}#1}}

\newcommand{\de}{~\cite{Carroll:1998zi,Fujita:2020aqt,Panda:2010uq,Fujita:2020ecn,Choi:2021aze,Obata:2021nql,Gasparotto:2022uqo,Galaverni:2023zhv}~}
\newcommand{\ede}{~\cite{Murai:2022zur,Eskilt:2023nxm,Fujita:2020ecn}~}
\newcommand{\dm}{~\cite{Finelli:2008jv,Liu:2016dcg,Fedderke:2019ajk}~}
\newcommand{\td}{~\cite{Takahashi:2020tqv,Kitajima:2022jzz,Jain:2022jrp,Gonzalez:2022mcx}~}
\newcommand{\qg}{~\cite{Myers:2003fd,Balaji:2003sw,Arvanitaki:2009fg}~}

\newcommand{\ebtomo}{~\cite{Sigl:2018fba,Sherwin:2021vgb,Nakatsuka:2022epj,Liu:2006uh,Lee:2013mqa,Gubitosi:2014cua,Eskilt:2023nxm,Fujita:2020ecn,Finelli:2008jv}}
\newcommand{\ebtomoinst}{~\cite{QUaD:2008ado,Komatsu:2011a,BICEP1:2013rur,Planck:2016soo,Monelli:2022pru}~}

\abstract{
We present a study on using delensing to enhance cosmic birefringence measurements based on full-sky, map-based simulations. In our analysis, we neglect foreground contamination and instrumental systematics to isolate the intrinsic impact of delensing on both isotropic and anisotropic birefringence. For the isotropic case, assuming a constant rotation angle of $\beta = 0.35^\circ$, delensing reduces the lensing-induced variance in the $EB$ power spectrum, yielding an improvement in sensitivity of approximately 10\% at 6$\mu$K-arcmin noise for $\ell < 200$ and 25–40\% at lower noise levels for $\ell < 1000$. For the anisotropic case, using simulations at 1 $\mu$K-arcmin noise, we reconstruct the birefringence angle for a scale-invariant spectrum and mitigate lensing bias by delensing, achieving a 50\% reduction in the leading $N_{(0)}$ bias and a 30\% improvement in the constraints on the amplitude $A_{\rm CB}$. Our results demonstrate that delensing is an effective tool for enhancing the detectability of subtle parity-violating signals in the CMB with forthcoming experiments such as the Simons Observatory, CMB-S4, and LiteBIRD.}

\arxivnumber{2503.04708}
\title{Improving Cosmic Birefringence Constraints via Delensing}

\author[\orcidlink{0000-0003-1200-9179}]{Anto Idicherian Lonappan}
\affiliation{Department of Physics, \\
University of California.\\
9500 Gilman Dr. \\
San Diego, CA 92093, USA.}

\emailAdd{alonappan@ucsd.edu}

\begin{document}
\maketitle
\flushbottom


\section{Introduction} \label{sec:intro}
The cosmic microwave background (CMB) serves as a potent observational tool for probing the early universe, encoding invaluable information about fundamental physics. Among the diverse phenomena accessible through CMB polarization, cosmic birefringence\textemdash the rotation of the linear polarization plane of CMB photons as they traverse cosmological distances\textemdash has recently attracted considerable attention as a potential indicator of new physics beyond the Standard Model~\citep{Carroll:1989vb,Carroll:1998zi,Carroll:1991zs,Harari:1992ea,Ni:2004jj,Ni:2007ar,Kostelecky:2007zz}. A robust detection of cosmic birefringence would provide strong evidence for parity-violating interactions in the universe.

Cosmic birefringence (CB) is broadly classified into isotropic and anisotropic categories. In the isotropic case, the CMB polarization undergoes a uniform rotation across the sky, potentially signaling a isotropic pseudoscalar field interacting with photons, which may approximate the behavior of dark energy \de. Conversely, anisotropic CB exhibits a position-dependent rotation angle, driven by fluctuations in the pseudoscalar field. Various cosmological phenomena—including dark matter \dm, early dark energy \ede, dark energy\de, and topological defects \td—have been proposed as sources of axion-like particles(ALP) that could induce CB via Chern-Simons coupling~\citep{Ni:2004jj}. If ALPs are responsible for CB, this phenomenon provides a robust indirect probe of their properties, including mass, coupling strength, and cosmic evolution. Furthermore, the temporal evolution of pseudoscalar fields during recombination and reionization can alter the $EB$ power spectrum, affecting both its amplitude and shape. Precise reconstruction of this spectral structure offers a tomographic window into the dynamics of pseudoscalar fields throughout cosmic history\ebtomo, while also mitigating degeneracies arising from instrumental systematics\ebtomoinst. Additionally, CB may reveal observable signatures of parity-violating physics predicted by quantum gravity theories\qg.

\revr{Recent analyses, including measurements from \textit{Planck}~\citep{Sullivan:2025btc,Minami:2020odp,Diego-Palazuelos:2022dsq,Diego-Palazuelos:2022cnh,Eskilt:2022wav,Planck:2016soo,Eskilt:2022cff}, ACT~\citep{ACT:2025fju}, and joint analyses of \textit{Planck} and WMAP~\citep{Cosmoglobe:2023pgf}, indicate a non-zero isotropic birefringence angle. There are also measurements from QUaD~\citep{QUaD:2008ado}, WMAP~\citep{2013ApJS20819H}, POLARBEAR~\citep{POLARBEAR:2019kzz}, SPT~\citep{SPT:2020cxx}, and BICEP~\citep{BICEPKeck:2024cmk} that have explored the isotropic birefringence angle.} However, these findings require further confirmation through upcoming experiments such as the \revr{Simons Observatory~\citep{so,Jost:2022oab}, LiteBIRD~\citep{lb,LiteBIRD:2025mvy},} CMB-S4~\citep{s4}, \revr{AliCPT~\cite{Zhong:2024tgw,Dou:2024wdy}} and CMB-HD~\citep{CMB-HD:2022bsz}. For anisotropic CB, the most stringent constraint on the amplitude of the scale-invariant rotation spectrum is \revr{$A_{\mathrm{CB}} = L(L+1)C_L^{\alpha\alpha}/2\pi \leq 0.044 \times 10^{-4}\;[\rm rad^2]$} at the 95\% confidence level~\citep{BICEPKeck:2022kci}; additional constraints based on \textit{Planck} data have also been reported~\citep{Gruppuso:2020kfy,Bortolami:2022whx,Zagatti:2024jxm}. Future observations are expected to enhance these constraints even further, improving our sensitivity to spatial fluctuations in the rotation angle.

However, a recent study has shown there is a measurable gravitational lensing impact on CB~\citep{Naokawa:2023upt}. Gravitational lensing alters CMB polarization by deflecting photon trajectories through large-scale structures, which converts $E$-modes into $B$-modes and increases the overall variance~\citep{Zaldarriaga:1998ar}. In this work, we build on prior work by applying delensing techniques\textemdash developed to tighten constraints on primordial gravitational waves~\citep{Kesden:2002ku,Seljak:2003pn,smith:2012a}, sharpen acoustic peaks~\citep{Planck:2018lbu}, improve cosmological parameter estimation~\citep{Green:2016cjr,Hotinli:2021umk}, probe primordial non-Gaussianity~\citep{Coulton:2019odk}\textemdash to measure CB~\citep{BICEPKeck:2024cmk} with enhanced precision. Our analysis focuses on reducing lensing-induced variance to refine constraints on both isotropic and anisotropic birefringence, systematically assessing potential biases and quantifying improvements in statistical sensitivity.

This paper is structured as follows: In Section \ref{sec:theory}, we outline the theoretical framework of CB and its connection to new physics. Section \ref{sec:sim} describes our simulation methodology, detailing the modeling of CMB polarization, lensing, and birefringence effects. In Section \ref{sec:qe}, we present the quadratic estimation techniques used for both lensing and birefringence reconstruction. Section \ref{sec:delens} focuses on delensing, including its impact on isotropic and anisotropic birefringence measurements, as well as the possible biases introduced. Finally, in Section \ref{sec:conclusion}, we discuss the broader implications of our findings and future directions for improving CB constraints in upcoming CMB experiments.\\
\section{Theoretical Framework} \label{sec:theory}
In this section, we establish the theoretical framework essential for understanding the influence of gravitational lensing and CB on CMB polarization. We first examine weak gravitational lensing, which distorts the polarization pattern and generates B-modes, thereby introducing an additional source of noise into birefringence measurements. We then delineate the physics underlying CB, distinguishing between its isotropic and anisotropic manifestations. Our analysis is focused on how these phenomena modify the polarization of the CMB.

\subsection{Weak Gravitational Lensing of CMB}
As CMB photons traverse the Universe, their trajectories are deflected by the gravitational potentials associated with intervening large-scale structures. This phenomenon, known as \textit{gravitational lensing}, distorts the observed CMB anisotropies and modifies the intrinsic polarization patterns~\citep{Challinor:2002cd,Lewis:2006fu,hanson:2010}. In particular, gravitational lensing remaps the Stokes parameters $Q'$ and $U'$, which describe the linear polarization of the CMB, according to
\begin{equation} 
\tilde{Q} \pm i \tilde{U} (\hat{n}) = [Q' \pm iU'] (\hat{n} + \mathbf{d}), 
\end{equation}
where $\hat{n}$ is the unit vector along the line of sight and $\mathbf{d}$ is the deflection vector. The deflection vector can be decomposed into a gradient (curl-free) component and a curl component:
\begin{equation} \mathbf{d} = \nabla \phi + \nabla \times \psi, \end{equation}
where $\phi$ is the lensing potential and $\psi$ represents the curl mode. Under the standard assumptions of weak lensing and the Born approximation, the curl component is expected to vanish {($\psi \approx 0$)~\citep{Namikawa:2011cs,Robertson:2024qjl},} so that the deflection is well approximated by
\begin{equation} 
\mathbf{d} \approx \nabla \phi. 
\end{equation}
The lensing potential $\phi$ is given by the line-of-sight projection of the three-dimensional gravitational potential (also referred to as the Weyl potential $\Psi$~\citep{Lewis:2006fu}), which is sourced by the matter distribution:
\begin{equation} 
\phi (\hat{n}) = -2 \int_0^{\chi_*} d\chi \frac{\chi_* - \chi}{\chi_* \chi} \Psi (\chi \hat{n}, \eta_0 - \chi), 
\end{equation}
where $\chi$ is the conformal distance, $\eta_0$ is the conformal time today, and $\chi_*$ is the conformal distance to the last-scattering surface. The gravitational potential $\Psi (\chi \hat{n}, \eta_0 - \chi)$ is evaluated along the unperturbed photon trajectory, consistent with the Born approximation, which assumes that photons travel in straight lines prior to being lensed~\citep{Pratten:2016dsm,Lewis:2016tuj,Fabbian:2017wfp}.

One of the key observational consequences of CMB lensing is the conversion of $E$-mode polarization to $B$-mode polarization. The lensing-induced $B$-mode power spectrum introduces additional noise, comparable to a 5 $\mu$K-arcmin white noise up to degree scales~\citep{Zaldarriaga:1998ar}. This lensing $B$-mode signal has been detected by CMB experiments and poses a significant challenge for detecting primordial gravitational waves~\citep{POLARBEAR:2022dxa,BICEPKeck:2022mhb,SPIDER:2021ncy,SPTpol:2020rqg,SPT:2019nip,planck2018lensing}. Understanding and mitigating the effects of weak gravitational lensing is crucial for improving the precision of cosmological measurements, including constraints on how CB affects the polarization of the CMB.\\

\subsection{Cosmic Birefringence}
CB arises when CMB photons interact with a pseudoscalar field, such as ALPs, which couples to electromagnetism through a Chern-Simons term~\citep{Carroll:1989vb,Carroll:1991zs,Komatsu:2022nvu}. The presence of CB introduces parity-violating signatures in the CMB, specifically generating nonzero $TB$ and $EB$ correlations that are absent in the standard $\Lambda$CDM model~\citep{Lue:1998mq,Feng:2004mq,Liu:2006uh}.

The interaction responsible for CB is described by the following Lagrangian density:
\begin{equation}
\mathcal{L} \supset -\frac{1}{4} F^{\mu\nu} F_{\mu\nu} - \frac{1}{4} g a F^{\mu\nu} \tilde{F}_{\mu\nu},
\end{equation}
where $F^{\mu\nu}$ is the electromagnetic field strength tensor, $\tilde{F}^{\mu\nu}$ is its dual, $a$ is the pseudoscalar field (such as an ALP), and $g$ is the axion-photon coupling constant. This interaction results in a rotation of the linear polarization angle of CMB photons as they traverse space filled with a nontrivial $a$ field. The total birefringence rotation angle is given by:
\begin{equation}
{\beta = \frac{1}{2} g \int_{t_{\mathrm{LSS}}}^{t_0} dt \frac{da}{dt},}
\end{equation}
where $t_{\mathrm{LSS}}$ denotes the time of last scattering and $t_0$ is the present time. This equation shows that the observed birefringence angle depends on the evolution of the $a$ field along the photon’s trajectory.

Studying CB offers a unique avenue to probe physics beyond the Standard Model, including the nature of dark energy, axionic dark matter, and potential violations of fundamental symmetries. In this work, we consider two primary manifestations of CB. The first, isotropic CB, examines a uniform rotation of the CMB polarization plane across the entire sky, while the second, anisotropic CB, accounts for spatial variations in the rotation angle. These variations provide critical insights into the distribution and evolution of the underlying pseudoscalar field. The following subsections provide a brief exploration of these cases.
\subsubsection{Isotropic case}
For the isotropic case, where the birefringence angle, $\beta$ is uniform across the sky, the transformation of the Stokes parameters $Q$ and $U$ due to a rotation can be expressed as follows:
\begin{equation}
Q' \pm iU' = (Q \pm iU) e^{\pm 2i\beta}.
\end{equation}
Here, $Q'$ and $U'$ denote the observed Stokes parameters after rotation, while 
$Q$ and $U$ represent the intrinsic polarization before the effect of birefringence. Thus, the polarization components transform as:
\begin{equation}
\begin{pmatrix} E'_{\ell m} \\ B'_{\ell m} \end{pmatrix} =
\begin{pmatrix} \cos(2\beta) & -\sin(2\beta) \\ \sin(2\beta) & \cos(2\beta) \end{pmatrix}
\begin{pmatrix} E_{\ell m} \\ B_{\ell m} \end{pmatrix}.
\end{equation}
which leads to modifications in the observed $EB$ angular power spectra~\citep{Lue:1998mq,Feng:2004mq,Liu:2006uh}:
\begin{equation}\label{eq:EB}
C^{E'B'}_\ell = \frac{1}{2} \sin(4\beta) \left(C^{EE}_\ell - C^{BB}_\ell \right).
\end{equation}
{This model assumes an ultralight pseudoscalar field with mass $m \lesssim 10^{-28}~\mathrm{eV}$, for which the rotation evolves slowly over cosmic time. For heavier fields, the $C_\ell^{EB}$ spectrum may exhibit a more complex scale dependence.}
\subsubsection{Anisotropic case}

The modified Stokes parameters in the presence of anisotropic birefringence are given by:
\begin{equation}
Q' (\hat{n}) \pm iU' (\hat{n}) = \left[ Q (\hat{n}) \pm iU (\hat{n}) \right] e^{\pm 2i \alpha(\hat{n})}.
\end{equation}
Here, $\alpha(\hat{n})$\footnote{{We use $\alpha$ for the anisotropic birefringence angle to distinguish it from the isotropic angle $\beta$.      }} represents the spatially varying birefringence angle, which describes the rotation of the polarization plane at different positions on the sky~\citep{vera:2009}. This angle is a scalar field that can be expanded in terms of spherical harmonics:
\begin{equation}
\alpha(\hat{n}) = \sum_{LM} \alpha_{LM} Y_{LM}(\hat{n}),
\end{equation}
where $\alpha_{LM}$ are the expansion coefficients and $Y_{LM}(\hat{n})$ are the spherical harmonics. Unlike the isotropic case, where the rotation angle is constant, anisotropic birefringence introduces spatially dependent modifications to the CMB polarization. This spatial variation leads to additional mode couplings in the observed CMB power spectra, encoding information about the underlying physics driving the birefringence effect.

\revr{The effect of anisotropic birefringence on the CMB polarization induces mode coupling between $E$- and $B$-modes, modifying them as follows \cite{Namikawa:2020ffr}:

\begin{equation}
E'_{\ell m} \pm i B'_{\ell m} = E_{\ell m} \pm i B_{\ell m} + \sum_{LM \ell' m'} (-1)^m 
\begin{pmatrix} \ell & L & \ell' \\ -m & M & m' \end{pmatrix}
W^{\pm}_{\ell L \ell'} [E_{\ell' m'} \pm i B_{\ell' m'}] \alpha_{LM},
\end{equation}
where the coupling coefficient $W^{\pm}_{\ell L \ell'}$ is defined as:
\begin{equation}
W^{\pm}_{\ell L \ell'} = \pm 2 \zeta^{\mp} p^{\pm}_{\ell L \ell'} 
\sqrt{\frac{(2\ell + 1)(2L + 1)(2\ell' + 1)}{4\pi}} 
\begin{pmatrix}
\ell & L & \ell' \\
-2 & 0 & 2
\end{pmatrix}.
\end{equation}
Where $\zeta^+ = 1$, $\zeta^- = i$ and the parity indicator is given by:
\begin{equation}
p^{\pm}_{\ell L \ell'} = \frac{1 \pm (-1)^{\ell + L + \ell'}}{2}\,.
\end{equation}
The terms in parentheses represent Wigner 3j symbols. This formulation illustrates that anisotropic birefringence introduces mode coupling across different multipoles, resulting in non-zero off-diagonal elements in the covariance matrix.}

\section{Simulations} \label{sec:sim}
In this study, we conduct a full-sky, map-based delensing analysis to investigate the effects of gravitational lensing and delensing on CB, utilizing CMB polarization maps. These maps are generated using the HEALPix pixelization scheme with a resolution parameter of $N_{\rm side} = 2048$. We assess three distinct noise levels—6, 2, and 1 $\mu$K arcmin—reflecting the expected sensitivities of upcoming CMB experiments such as the Simons Observatory, LiteBIRD and CMB-S4, respectively, while keeping the experimental beam fixed at 1 arcmin across all cases.

To produce the lensed CMB polarization maps, we start with unlensed CMB polarization maps and a lensing potential map, both generated as random Gaussian fields\footnote{We note that our simulations neglect the correlation between the unlensed CMB anisotropies and the lensing potential. This choice is justified in the context of our analysis, as we do not model the higher-order lensing effects where such correlations become relevant.} based on the fiducial angular power spectra computed using \texttt{CAMB}\footnote{\url{https://github.com/cmbant/camb}}~\cite{Lewis:1999bs}. Gravitational lensing is then applied using the \texttt{lenspyx}\footnote{\url{https://github.com/carronj/lenspyx}}~\citep{Reinecke:2023gtp} package, which employs bicubic interpolation on an oversampled equidistant cylindrical projection grid to ensure high precision. We explore two scenarios of CB:
\begin{enumerate}
  \item \textbf{Isotropic Birefringence}: An isotropic rotation angle of $ \beta = 0.35^\circ$ is applied to the unlensed CMB polarization maps before lensing.
  \item \textbf{Anisotropic Birefringence}: The birefringence field is modeled with a scale-invariant power spectrum given by
\begin{equation}\label{eq:aniso_spectrum}
    C_{L}^{\alpha\alpha} = A_{\rm CB} \frac{2\pi}{L(L+1)},
\end{equation}
    where the amplitude is set to $A_{\rm CB} = 1 \times 10^{-6}$. Gaussian realizations of the rotation field $\alpha$ are applied to the unlensed maps prior to lensing. Although observational upper limits on $A_{\rm CB}$ are as high as $0.044\times 10^{-4}$, we adopt this smaller value to probe subtle birefringence signals. {We adopt a scale-invariant form for $C_L^{\alpha\alpha}$ as a minimal and widely used template, particularly relevant for light pseudoscalar fields. Other mechanisms, such as topological defects may produce non–scale-invariant spectra, which will be explored in future work.}
\end{enumerate}

\begin{figure}
    \centering
   \includegraphics[width=\linewidth]{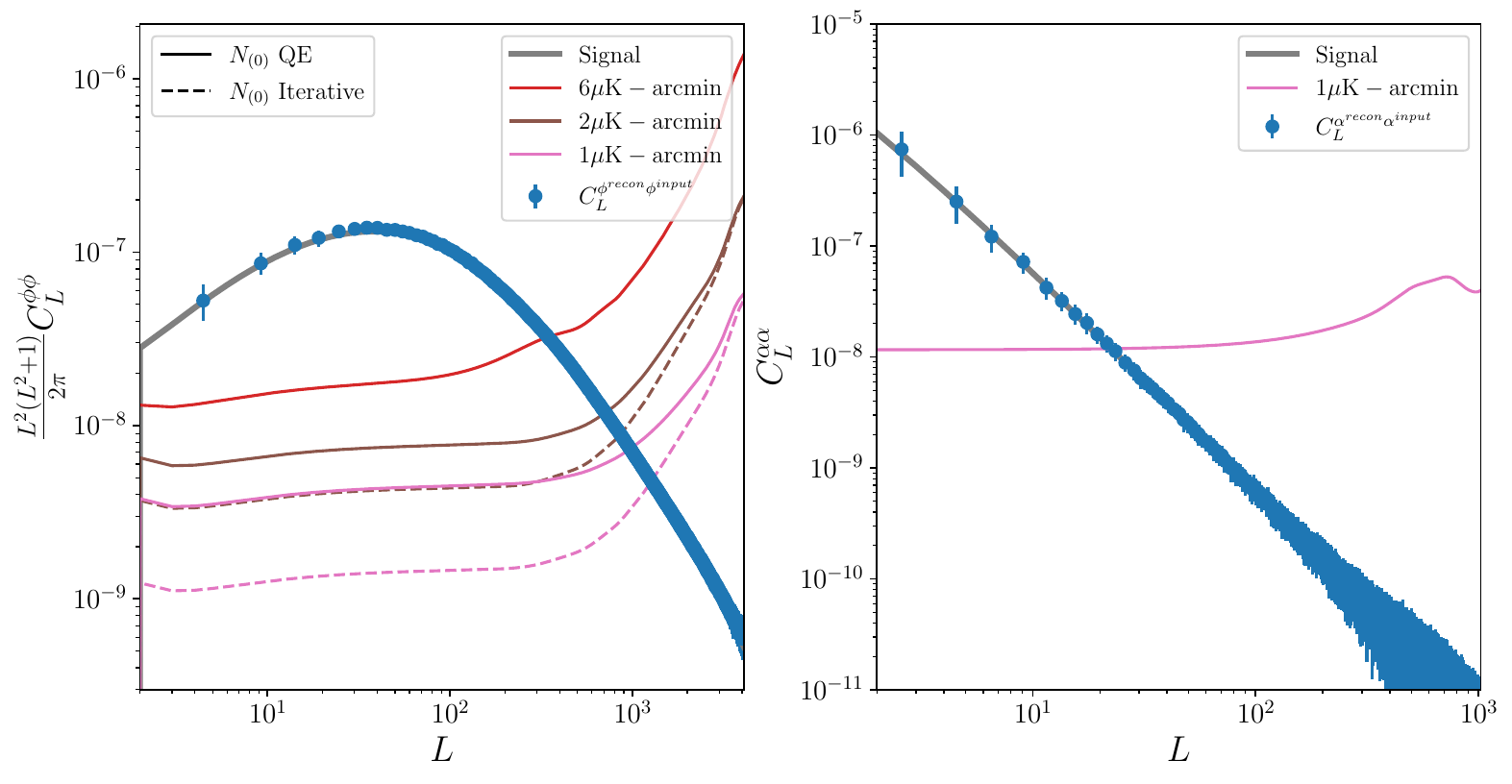}
   \caption{{\textbf{Left}: Binned reconstructed lensing potential power spectrum cross-correlated with the input $\phi_{LM}$. \textbf{Right}: Binned power spectrum of the reconstructed anisotropic birefringence angle cross-correlated with the input $\alpha_{LM}$. In both panels, the blue points with error bars represent the binned reconstructed spectra, and the grey solid line indicates the input signal. The error bars denote the standard deviation from 100 simulations. In the left panel, the solid colored lines show the $N^\phi_{(0)}$ bias and the dashed colored lines depict the iterative $N^\phi_{(0)}$ bias, with red, brown, and pink corresponding to noise levels of 6, 2, and 1 $\mu$K-arcmin, respectively. In the right panel, the solid pink line denotes the $N^\alpha_{(0)}$ bias for the 1 $\mu$K-arcmin case.}}
    \label{fig:qe}
\end{figure}

\section{Quadratic Estimation}\label{sec:qe}
In this section, we outline our methodology for reconstructing both the lensing potential and the birefringence angle via quadratic estimators. These estimators enable the extraction of the statistical anisotropies introduced by gravitational lensing and CB from CMB polarization maps. We employed quadratic estimators implemented in the \texttt{plancklens} package\footnote{\url{https://github.com/carronj/plancklens/}}~\citep{Planck:2018lbu}. Prior to reconstruction, we apply Wiener filtering to optimally extract the relevant signals while suppressing noise~\citep{hanson:2009u}. The Wiener-filtered fields are given by:
\begin{equation}
X^{\text{WF}}_{\ell m} = \frac{\revr{C_{\ell}^{XX}}}{C_{\ell}^{XX} + N_{\ell}^{XX}} X_{\ell m},
\end{equation}
where $X \in [E,B]$, $C_{\ell}^{XX}$ is the theoretical power spectrum, and $N_{\ell}^{XX}$ is the noise power spectrum.

\subsection{Lensing Reconstruction}
The off-diagonal elements of the covariance, resulting from anisotropies in the deflection field~\citep{Okamoto:2003zw,Namikawa:2014yca}, are expressed as:
\begin{equation}
\langle E_{\ell m} B_{\ell' m'} \rangle_{(\ell m)\neq (\ell' -m')} = \sum_{LM} (-1)^M
\begin{pmatrix} \ell & \ell' & L \\ m & m' & -M \end{pmatrix}
f^{EB}_{\ell L \ell'} \phi_{LM},
\end{equation}
where $\phi_{LM}$ denotes the lensing potential {and parentheses represents the Wigner's 3j symbol.} The $f^{EB}_{\ell L \ell'}$ is the weighting function that quantifies the response of the observed CMB polarization to the mode coupling induced by gravitational lensing. This function characterizes how off-diagonal correlations are affected by lensing, with its explicit form provided in Table 1 of Ref. \citep{Okamoto:2003zw}:\revr{
\begin{equation}
f^{EB}_{\ell L \ell'} = -i p^{-}_{\ell L \ell'} 
\sqrt{\frac{(2\ell + 1)(2L + 1)(2\ell' + 1)}{16\pi}} 
\left[ L(L+1) + \ell(\ell+1) - \ell'(\ell'+1) \right]
\begin{pmatrix}
\ell' & L & \ell \\
2 & 0 & -2
\end{pmatrix}
C^{EE}_{\ell},
\end{equation}
where $C^{EE}_{\ell}$ is the $E$-mode power spectrum. Note that in $f^{EB}_{\ell L \ell'}$, we have omitted the negligible contribution from $B$-modes. A key property of gravitational lensing is that it conserves parity, implying that $f^{EB}_{\ell L \ell'}$ is non-zero only when the sum $\ell + L + \ell'$ is odd. Although gravitational lensing converts $E$-modes into $B$-modes, it does not induce an ensemble-averaged $EB$ correlation in the CMB. Thus, the lensing potential can be reconstructed using} a quadratic estimator of the CMB anisotropies, and the unnormalized estimator is defined as:
\begin{equation}
\tilde{\phi}_{LM} = \sum_{\ell \ell'}
\begin{pmatrix} \ell & \ell' & L \\ m & m' & M \end{pmatrix}
(f^{EB}_{\ell L \ell'})^{} E_{\ell m} B_{\ell' m'}.
\end{equation}
The normalization in the idealized full-sky case is given by:
\begin{equation}
A^{\phi}_L = \left[ \frac{1}{2L+1} \sum_{\ell \ell'} \frac{|f^{EB}_{\ell L \ell'}|^2}{C^{EE}_\ell C^{BB}_\ell} \right]^{-1}.
\end{equation}
The reconstruction of the lensing potential, up to first order in $C_\ell^{\phi\phi}$, is governed by two primary biases. The first, {$N_{(0)}^{\phi}$}, arises from the disconnected part of the four-point correlation function, while the second, {$N_{(1)}^{\phi}$}, originates from the secondary contraction of the connected components at first order in the lensing potential,though its amplitude remains typically low compared to {$N_{(0)}^{\phi}$}. We follow the methodology outlined in Ref. \citep{planck2018lensing} to estimate these biases. While both {$N_{(0)}^{\phi}$} and {$N_{(1)}^{\phi}$} biases are present, {the Wiener filtering uses the $N_{(0)}^{\phi}$ bias which is equivalent to $A_L^{\phi}$}(see Section \ref{sec:delens} for details). The left panel of Fig. \ref{fig:qe} displays the binned reconstructed lensing potential power spectrum, cross-correlated with the input $\phi_{LM}$. The {$N_{(0)}^{\phi}$} bias for noise levels of 6, 2, and 1 $\mu$K-arcmin is shown in red, brown, and pink, respectively. For low-noise experiments, the quadratic {$N_{(0)}^{\phi}$} estimator becomes suboptimal, and we can improve the estimate using an iterative lensing approach~\citep{smith:2012kmh,Seljak:2003pn}. The same figure also shows the iterative {$N_{(0)}^{\phi}$} bias, estimated using the \texttt{cmblensplus}\footnote{\url{https://github.com/toshiyan/cmblensplus}}~\citep{Namikawa:2020ffr} package, represented by dashed lines. {In this analysis, we use the reconstructed lensing potential to delens the polarization field. The lensing potential is reconstructed using CMB polarization fields over the multipole range $2 \leq \ell \leq 4096$, and we reconstruct the lensing potential up to $\ell=4096$.}

\begin{figure}
    \centering
    \begin{subfigure}{0.45\textwidth}
        \centering
        \includegraphics[width=\linewidth]{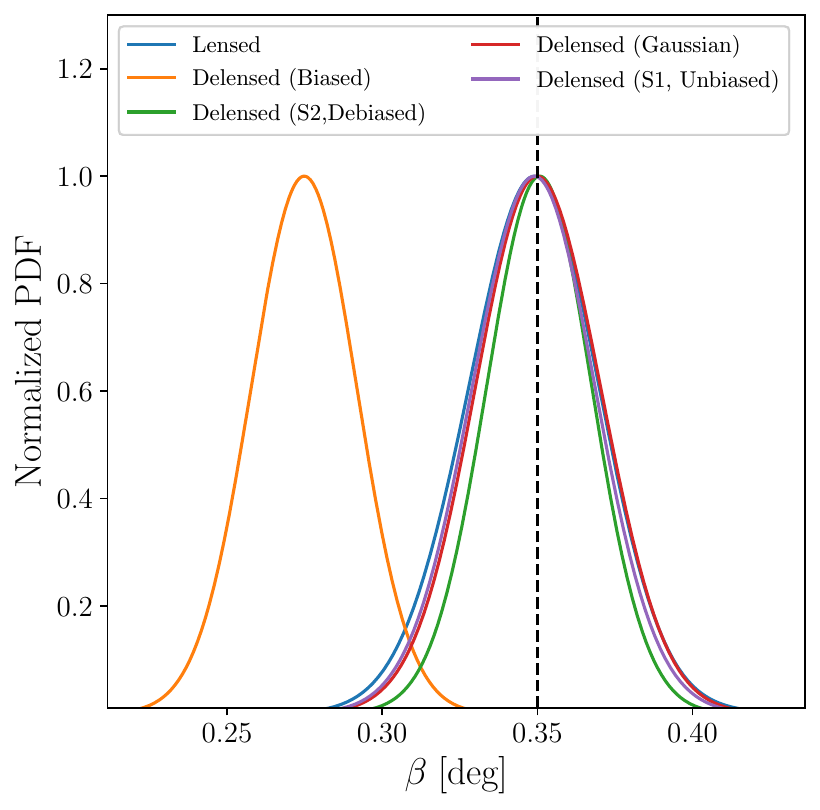}
    \end{subfigure}
    \begin{subfigure}{0.45\textwidth}
        \centering
        \includegraphics[width=\linewidth]{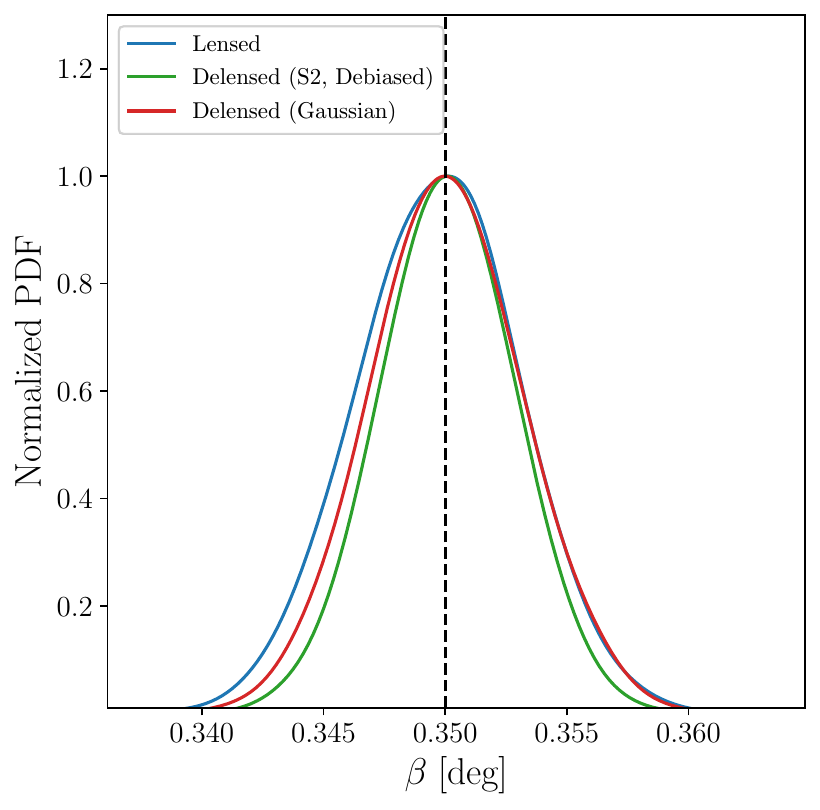}
    \end{subfigure}
\caption{Posterior distributions of the isotropic birefringence angle $\beta$. \textbf{Left} Posterior obtained over the multipole range $2 \leq \ell \leq 200$. {The lensed posterior is shown in blue, while the orange line indicates the impact of internal delensing bias. The green line represents the debiased case using the S2 strategy, and the purple line shows the S1-based unbiased estimate. We also show the posterior obtained using Gaussian $N^\phi_{(0)}$ in red. \textbf{Right:} Posterior distribution obtained over the multipole range $2 \leq \ell \leq 1000$, with the lensed posterior in blue and the delensed (S2) case in green and the red denotes the case with the Gaussian noise bias. The mean values and 68\% confidence levels of these posterior distributions are summarized in Table \ref{tab:beta_values}.}}
    \label{fig:highEB}
\end{figure}

\subsection{Birefringence Angle Reconstruction}
The off-diagonal elements of the covariance, resulting from anisotropies in the rotation field~\citep{vera:2009,Namikawa:2020ffr}, are expressed as:
\begin{equation}
\langle E_{\ell m} B_{\ell' m'} \rangle_{(\ell m)\neq(\ell' -m')} = \sum_{LM} (-1)^M
\begin{pmatrix} \ell & \ell' & L \\ m & m' & -M \end{pmatrix}
f^{\alpha}_{\ell L \ell'} \alpha_{LM},
\end{equation}
where $\alpha_{LM}$ represents the birefringence rotation field, and $f^{\alpha}_{\ell L \ell'}$(see Ref \citep{vera:2009}) serves as the weighting function that determines the response of the observed CMB polarization to the mode coupling induced by birefringence. \revr{The function $f^{\alpha}_{\ell L \ell'} = -W^-_{\ell' L \ell} \, C_\ell^{EE}$ characterizes how off-diagonal correlations respond to the birefringence effect, analogous to the response function for lensing-induced mode couplings. In this expression, the contribution from $C_\ell^{BB}$ has been neglected due to its relatively small amplitude. Moreover, $f^\alpha_{\ell L \ell'}$ is non-zero only when $\ell + L + \ell'$ is even.} Consequently, the birefringence field can be reconstructed as a quadratic estimator of the CMB anisotropies, and the unnormalized estimator is defined as:
\begin{equation}
\tilde{\alpha}_{LM} = \sum_{\ell \ell'}
\begin{pmatrix} \ell & \ell' & L \\ m & m' & M \end{pmatrix}
(f^{\alpha}_{\ell L \ell'})^{} E_{\ell m} B_{\ell' m'}.
\end{equation}
The normalization in the idealized full-sky case is given by:
\begin{equation}\label{eq:norm_alpha}
A^{\alpha}_L =  \frac{1}{2L+1} \sum_{\ell \ell'} \frac{|f^{\alpha}_{\ell L \ell'}|^2}{C^{EE}_\ell C^{BB}_\ell}.
\end{equation}
Similar to the lensing reconstruction, the reconstruction of the birefringence angle is subject to a disconnected(Gaussian) bias, {with $N^\alpha_{(0)} = A_L^{\alpha}$ in the ideal case.} The right panel in Fig. \ref{fig:qe} presents the binned power spectrum of the reconstructed birefringence angle, cross-correlated with the input $\alpha_{LM}$. The {$N^\alpha_{(0)}$} bias is shown as a solid pink line for the 1 $\mu$K-arcmin noise simulation.
\section{Delensing}\label{sec:delens}
In this section, we aim to improve the sensitivity of CB measurements by reducing lensing-induced variance in CMB polarization. We implement \textit{internal delensing} using a template-based approach similar to Refs.~\citep{Sehgal:2016eag,Namikawa:2014yca,planck2018lensing}, reconstructing the lensing potential from the same CMB fields used in the analysis. Specifically, we filter the reconstructed lensing potential over the multipole range $8 \leq \ell \leq 4096$.
and compute a Wiener-filtered lensing potential, $\phi_{LM}^{WF}$, to optimally capture lensing modes while minimizing noise:
\begin{equation} \revr{\phi_{LM}^{WF} = \sqrt{L(L+1)} \frac{C_{L}^{\phi\phi}}{C_{L}^{\phi\phi} + N_{L}^{\phi\phi}} \phi_{LM}}, \end{equation}
where $C_{L}^{\phi\phi}$ is the theoretical lensing potential power spectrum, and $N_{L}^{\phi\phi}$ includes only the disconnected ($N_{(0)}$) bias in this work.
Next, we generate a Wiener-filtered $E$-mode map, $P^{(E)}(\hat{n}) = Q^{(E)}(\hat{n}) + \revr{i}U^{(E)}(\hat{n})$, and remap it according to $\nabla\phi_{LM}^{WF}$ to form a lensing B-mode template, $B_{\ell m}^{\mathrm{temp}}$. {Here, $P^{(E)}(\hat{n})$ denotes the real-space polarization field constructed from the $E$-mode map alone.} Finally, we subtract this template from the observed $B$-mode map, $B_{\ell m}$, to obtain the delensed $B$-mode map:
\begin{equation} B_{\ell m}^{\mathrm{del}} = B_{\ell m} - B_{\ell m}^{\mathrm{temp}}. \end{equation}
By removing these lensing-induced anisotropies, we enhance sensitivity to CB and refine parameter constraints. {We define the delensing efficiency as $\left(1 - \frac{A_{\mathrm{lens}}^{\mathrm{delensed}}}{A_{\mathrm{lens}}^{\mathrm{lensed}}} \right) \times 100$, where $A_{\mathrm{lens}}$ denotes the amplitude of the lensing $B$-mode power spectrum before and after delensing. The resulting efficiencies for different noise levels are summarized in Table~\ref{tab:delens_efficiency}.} The following subsections detail the delensing analysis for both isotropic and anisotropic CB.
\begin{table}[ht]
\centering
\begin{tabular}{cc}
\hline
Noise Level [$\mu$K-arcmin] & Delensing Efficiency [\%] \\
\hline
6                           & 16.01                      \\
2                           & 53.22 (59.06)              \\
1                           & 67.93 (76.68)              \\
\hline
\end{tabular}
\caption{{Delensing efficiencies achieved for different white-noise levels using the quadratic estimator. Values in parentheses indicate efficiencies obtained with the iterative $N^\phi_{(0)}$.}}
\label{tab:delens_efficiency}
\end{table}
\begin{figure}
    \centering
    \includegraphics[width=0.8\linewidth]{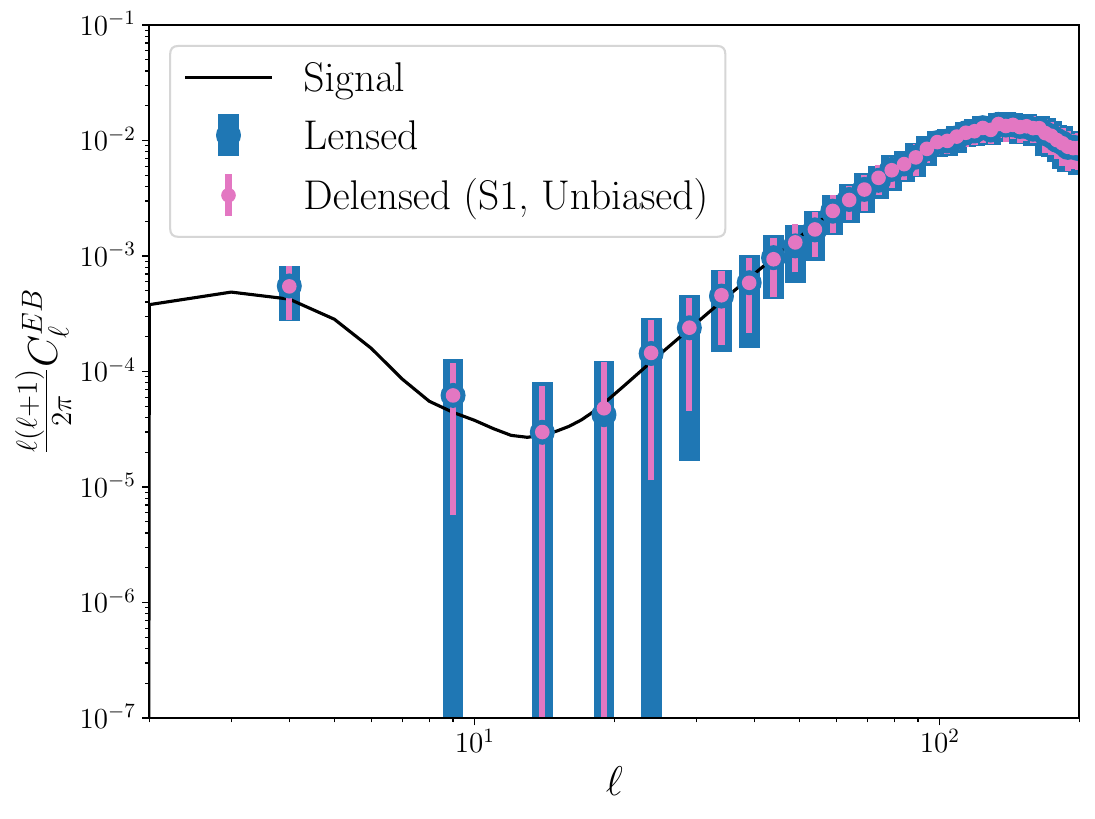}
    \caption{Binned $EB$ power spectrum at low $\ell$. The blue error bars represent the lensed spectrum, while the pink error bars correspond to the delensed spectrum with S1 strategy. The error bars are computed using 100 realizations.}
    \label{fig:lowEB}
\end{figure}

\subsection{Isotropic Cosmic Birefringence}
We estimate the $EB$ power spectra before and after the delensing procedure and assess the improvement in sensitivity using a likelihood approach. Before likelihood evaluation, the $C_\ell^{EB}$ spectra are binned linearly with a bin width of 5 multipoles, starting from $\ell=2$. The likelihood function is given by
\begin{equation}\label{eq:likelihood} 
\mathcal{L} \propto \exp\left( -\frac{1}{2} \mathbf{\Delta}^T \mathbf{C}^{-1} \mathbf{\Delta} \right), 
\end{equation}
where
\begin{equation} 
\mathbf{\Delta} = C_b^{EB,\text{data}} - C_b^{EB,\text{theory}},
\end{equation}
and $\mathbf{C}$ is the covariance matrix of the binned $C_\ell^{EB,\text{data}}$. {The covariance matrices are computed from 100 independent simulations, each including CMB signal, instrumental noise, and beam effects. In this analysis, we consider only the diagonal elements of $\mathbf{C}$, as the off-diagonal elements after delensing remain noisy and do not converge reliably with the available number of simulations. A significantly larger number of realizations would be required to accurately model the full off-diagonal covariance after delensing.} The posterior distribution of the birefringence angle $\beta$ is explored by sampling the likelihood using an affine-invariant MCMC sampler implemented in \texttt{emcee}\footnote{\url{https://github.com/dfm/emcee}}. For our initial analysis, we use the simulation set with 6~$\mu$K-arcmin noise. {Figure~\ref{fig:highEB} shows the posterior distributions of $\beta$ obtained from the lensed and delensed spectra, and the corresponding mean values and $1\sigma$ intervals are summarized in Table~\ref{tab:beta_values}}. As evident, the delensed power spectra exhibit a bias. We address this internal delensing bias in the following subsection.
\begin{table}[ht]
\centering
\begin{tabular}{lcc}
\hline
\multicolumn{1}{c}{} & \multicolumn{2}{c}{$\beta$ [deg]} \\
\cline{2-3}
Samples & $\ell < 200$ & $\ell < 1000$ \\
\hline
Lensed & $0.348 \pm 1.9 \times 10^{-2}$ & $0.349 \pm 3.2 \times 10^{-3}$ \\
Delensed (S2 Debiased) & $0.350 \pm 1.4 \times 10^{-2}$ & $0.350 \pm 2.6 \times 10^{-3}$ \\
Delensed (Gaussian) & $0.350 \pm 1.7 \times 10^{-2}$ & $0.350 \pm 3.0 \times 10^{-3}$ \\
Delensed (S1, Unbiased) & $0.348 \pm 1.7 \times 10^{-2}$ & --- \\
\hline
\end{tabular}
\caption{{Posterior mean and $1\sigma$ confidence intervals for the isotropic birefringence angle $\beta$ in degrees, estimated from lensed and delensed $C_\ell^{EB}$ spectra. Results are shown for two multipole ranges: $\ell < 200$ and $\ell < 1000$.}}
\label{tab:beta_values}
\end{table}
\subsubsection{Impact of Delensing Bias}\label{sec:delens_bias}
Internal delensing introduces a bias, known as the internal delensing bias (IDB), which arises because the noise in the reconstructed potential map is correlated with the CMB map being delensed. {This bias is absent when the lensing potential is obtained from external tracers~\cite{2012JCAP06014S} (e.g., galaxies or the cosmic infrared background) or reconstructed using the temperature field.} To mitigate IDB, we adopt two strategies:\\

\noindent\textbf{Strategy 1 (S1):} To prevent IDB, we reconstruct the lensing potential using CMB multipoles in the range $200 \leq L \leq 4096$, thereby avoiding any overlap with the modes used in the $EB$ power spectrum. Consequently, we restrict our analysis to the delensed $EB$ spectrum at $\ell < 200$, which remains unaffected by IDB. The purple curve in Figure~\ref{fig:highEB} (left) represents the posterior obtained using this method. The S1 shows an improvement of around 10\%. \revr{Figure~\ref{fig:lowEB} shows the binned $EB$ spectra for both the lensed and the delensed cases using the S1 strategy.}\\

\noindent\textbf{Strategy 2 (S2):} Following Ref \citep{planck2018lensing}, we repeat the delensing procedure using the input $\phi_{LM}$ supplemented with a Gaussian realization of the {$N^\phi_{(0)}$} bias. This approach prevents noise correlations from contaminating the delensing process. We then compute the IDB spectrum as 
\begin{equation} C_b^{\text{IDB}} = \left\langle C_b^{\text{EB, delensed, recon}} - C_b^{\text{EB, delensed, Gaussian}} \right\rangle, \end{equation} 
and subtract it from the delensed spectra. We find that this \textit{Planck}-like debiasing method becomes unreliable in the low-noise regime due to better signal-to-noise in the CMB polarization fields. \revr{The green curve in Figure~\ref{fig:highEB} represent the S2-based debiased posteriors. These are compared with the red curve, which correspond to delensing using a Gaussian realization of the $N^\phi_{(0)}$ bias}. Notably, the S2 strategy yields a lower variance than even the idealized Gaussian case, indicating a potential underestimation of the variance. This suggests that the bias subtraction from the biased delensed spectra in the S2 method may inadvertently reintroduce signal into the final spectrum, leading to an artificially tight constraint. Moreover, including high multipoles (extending the range to $2 \leq \ell \leq 1000$) further enhances the overall sensitivity as shown in the Figure~\ref{fig:highEB} (right): the lensed spectrum alone demonstrates an 80\% improvement, with debiased delensing providing an additional $\sim$7\% enhancement.\\
\begin{figure}
    \centering
    \includegraphics[width=\linewidth]{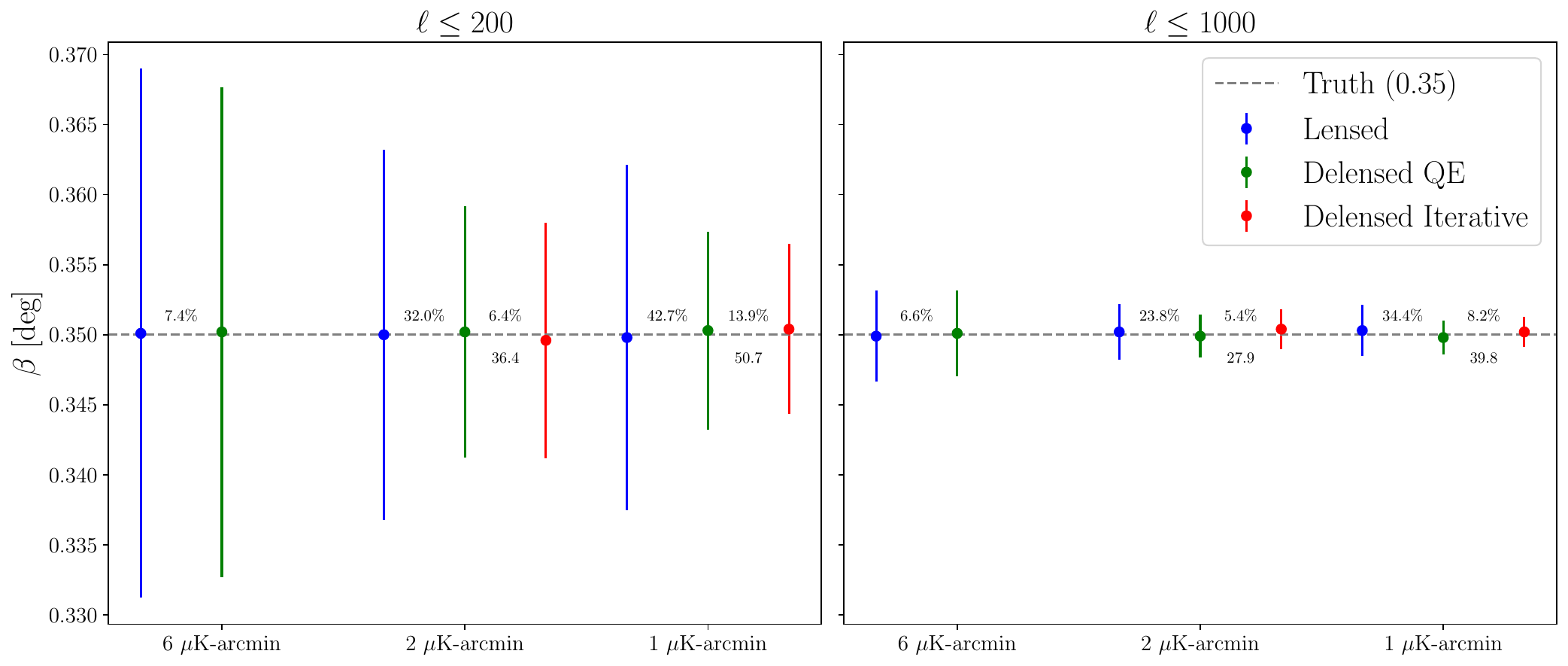}
\caption{Best-fit values of $\beta$ with corresponding 1$\sigma$ error bars for simulation sets at various noise levels. For the 6 $\mu$K-arcmin case, we show both the lensed and quadratic-estimator delensed results. For the lower noise levels (2 and 1 $\mu$K-arcmin), in addition to quadratic estimator delensing, we also present constraints from iterative delensing using a Gaussian iterative {$N^\phi_{(0)}$} approach. Blue error bars denote the lensed case, green error bars correspond to quadratic estimator delensed constraints, and red error bars indicate iterative {$N^\phi_{(0)}$} based constraints. The true value is marked by the dashed grey line. Sensitivity improvements from delensing are annotated above the dashed line, while percentage improvements comparing lensed and iterative delensing results are shown below it.}
    \label{fig:beta_noise}
\end{figure}
\subsubsection{Improvement by Iterative Lensing Estimates}
To further quantify the benefits of iterative delensing, we used the Gaussian simulations of iterative {$N^\phi_{(0)}$} bias estimates for noise levels of 2 $\mu$K-arcmin and 1 $\mu$K-arcmin. Figure~\ref{fig:beta_noise} presents the best-fit $\beta$ values with 1$\sigma$ confidence intervals for these noise levels. In this regime, delensing yields a sensitivity improvement of approximately 30\% for 2 $\mu$K-arcmin noise and 40\% for 1 $\mu$K-arcmin noise.
These results highlight the potential of internal delensing, combined with strategies to mitigate IDB, to significantly enhance the sensitivity of isotropic CB measurements in forthcoming CMB experiments.

\subsection{Anisotropic Cosmic Birefringence}
\begin{figure}
    \centering
    \begin{subfigure}{0.45\textwidth}
        \centering
        \includegraphics[width=\linewidth]{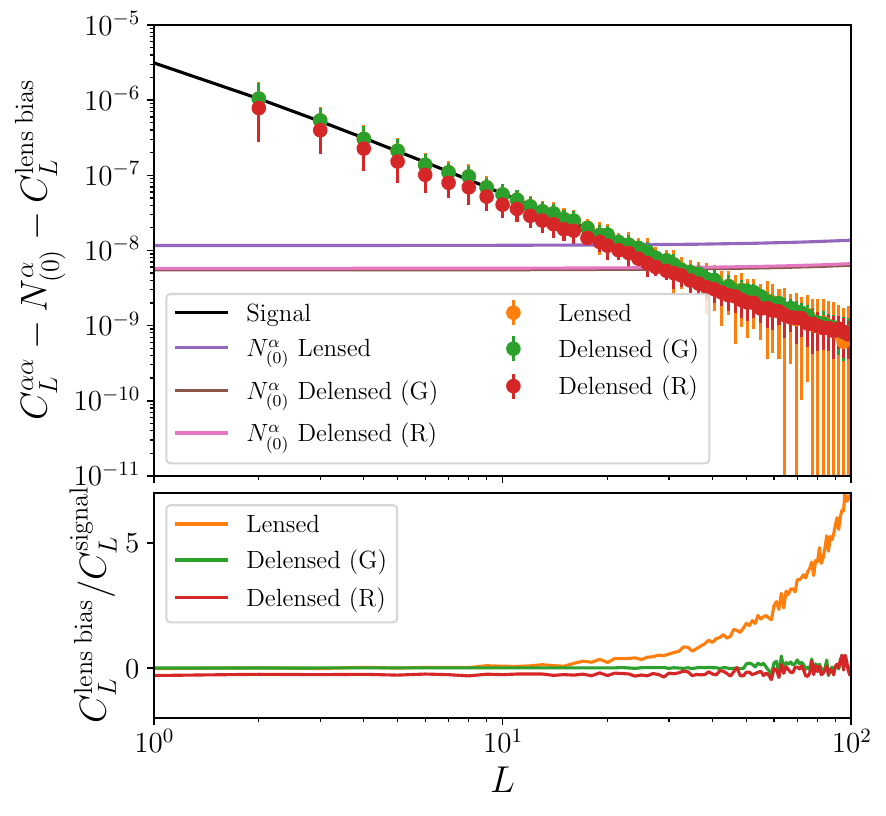}
    \end{subfigure}
    \begin{subfigure}{0.423\textwidth}
        \centering
        \includegraphics[width=\linewidth]{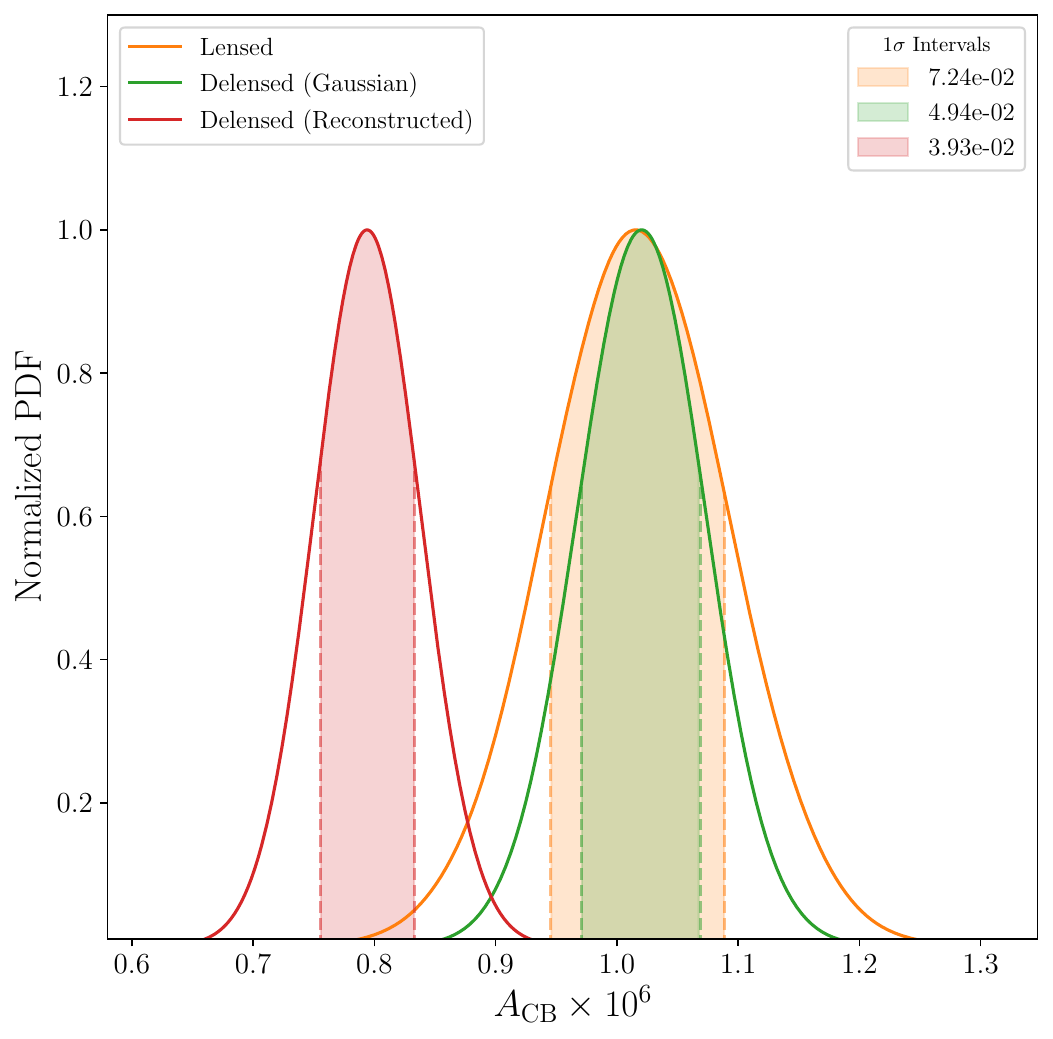}
    \end{subfigure}
\caption{{\textbf{Left (upper):} Binned reconstructed birefringence angle spectra after subtracting $N^\alpha_{(0)}$ and $C_L^{\rm lens\;bias}$. Orange points represent the lensed case (L), red points correspond to delensing using the reconstructed lensing potential (R), and green points correspond to delensing using the input (true) lensing potential with a Gaussian realization of $N^\phi_{(0)}$ (G). The purple, brown, and pink lines show the semi-analytical estimates of $N^\alpha_{(0)}$ for the L, G, and R cases, respectively. 
\textbf{Left (lower):} Bias-to-signal ratio for the same three cases. Orange, green, and red curves correspond to L, G, and R, respectively. 
\textbf{Right:} Posterior distribution of $A_{\rm CB}$. The  Orange, green, and red curves represent the L, G, and R cases, respectively. The shaded region indicates the 1$\sigma$ confidence interval, with values listed in the legend. All results are based on simulations with a noise level of 1~$\mu$K-arcmin; error bars reflect the standard deviation over 100 realizations.}}
    \label{fig:aniso}
\end{figure}

{For this analysis, we use a simulation set with a noise level of 1~$\mu$K-arcmin. We reconstruct the anisotropic CB angle using CMB polarization fields over the multipole range $2 \leq \ell \leq 1024$, thereby estimating the $\alpha_{LM}$ coefficients up to $L = 1024$. Gravitational lensing introduces significant biases in anisotropic birefringence estimates~\citep{Namikawa:2016fcz}, necessitating careful mitigation to enable robust parameter inference.
To quantify the impact of lensing-induced bias, we reconstruct the birefringence angle from two sets of simulations: one with standard lensed CMB maps, and another using Gaussian CMB simulations generated from the lensed CMB power spectra. The latter set retains the same statistical properties but lacks the non-Gaussian lensing-induced correlations, and is therefore expected to be free from lensing bias. We define the lensing bias spectrum as
\begin{equation}
C_L^\mathrm{lens\;bias} = \left\langle \left( C_L^{\alpha\alpha,\mathrm{recon}} - N^\alpha_{(0)} \right)_\mathrm{lensed\;CMB} - \left( C_L^{\alpha\alpha,\mathrm{recon}} - \revr{N^\alpha_{(0)}} \right)_\mathrm{Gaussian\;CMB} \right\rangle.
\end{equation}
Note that $C_L^{\mathrm{lens\;bias}}$ includes contributions from the lensing-induced component of $N_{(1)}^\alpha$.

The upper-left panel of Figure~\ref{fig:aniso} presents the bias-subtracted anisotropic CB power spectra. The orange points with error bars represent the spectrum reconstructed from lensed simulations. The red points show the delensed result using a lensing potential reconstructed from the CMB itself, while the green points correspond to delensing performed using the input (true) lensing potential with a Gaussian realization of $N^\phi_{(0)}$. The difference between these two delensed cases highlights the impact of IDB: a noticeable downward bias appears in the red points, which is absent in the green one. The lower panel shows the fractional bias, defined as the bias-to-signal ratio. The orange curve corresponds to the lensed case, where lensing-induced bias is clearly present. The green curve (delensed with input potential) shows effective removal of this bias, while the red curve (delensed with reconstructed potential) exhibits a residual suppression due to IDB.
We then perform a likelihood analysis—using the same formalism as for the isotropic case (see Eq.~\ref{eq:likelihood})\textemdash with}
\begin{equation}
\Delta = C_B^{\alpha\alpha,\mathrm{data}} - C_B^{\alpha\alpha,\mathrm{theory}}.
\end{equation}
{Here, $C_B^{\alpha\alpha,\mathrm{theory}}$ is given by Eq.~(\ref{eq:aniso_spectrum}). Before computing the likelihood, the reconstructed $C_L^{\alpha\alpha}$ spectra are binned into 200 intervals between $L=2$ and $L=1024$, using bin edges spaced uniformly in $\sqrt{L}$. This binning provides higher resolution at low multipoles, where the anisotropic birefringence signal is strongest, while maintaining adequate signal-to-noise at higher multipoles. Covariance matrices are estimated from 100 independent simulations. As in the isotropic case, we include only the diagonal elements of the covariance, since modeling the full off-diagonal terms would require accurate realization-dependent estimates of $N^\alpha_{(0)}$ and $N^\alpha_{(1)}$, especially after delensing.}
For the data spectrum, we define
\begin{equation}
C_B^{\alpha\alpha,\mathrm{data}} = C_B^{\alpha\alpha,\mathrm{recon}} - N^\alpha_{(0)} - C_B^\mathrm{lens\;bias},
\end{equation}
 Since we do not explicitly model or correct for the {$N^\alpha_{(1)}$} bias, our likelihood analysis is restricted to multipoles $\ell < 100$ to minimize its impact.
Our results indicate that the posterior on $A_{\rm CB}$ improves by approximately 30\% for the delensed case compared to the lensed case. This improvement is driven partly by the reduction in the $N^\alpha_{(0)}$ bias after delensing, but the dominant gain comes from suppressing lensing-induced bias. The right panel of Figure~\ref{fig:aniso} shows the posterior distributions of $A_{\rm CB}$ for the three cases: \revr{lensed (orange), delensed using the input lensing potential with Gaussian $N^\phi_{(0)}$ (green), and delensed using the reconstructed lensing potential (red).} These results demonstrate that delensing effectively suppresses both the leading-order {$N^\alpha_{(0)}$} and lensing bias, thereby enhancing the robustness and sensitivity of anisotropic CB measurements.

\section{Conclusion and Discussion}\label{sec:conclusion}
We have presented a comprehensive analysis of the impact of internal delensing on CB measurements using full-sky, map-based simulations. Conducted under idealized conditions—without foreground contamination or instrumental systematics—our study isolates the intrinsic improvements achievable through delensing and examines both isotropic and anisotropic birefringence scenarios.

For the isotropic case, we assumed a constant rotation angle of $\beta = 0.35^\circ$ and demonstrated that internal delensing substantially reduces the lensing-induced variance in the $EB$ power spectrum. Our likelihood analysis, based on the $EB$ spectrum, indicates an improvement in the sensitivity to $\beta$ of approximately 10\% for a noise level of 6 $\mu$K-arcmin, a sensitivity achievable by the Simons Observatory (both SAT and LAT). {In particular, to further mitigate delensing bias and achieve optimal performance, delensing using external tracers or temperature-based lensing reconstruction offers promising bias-free alternatives for upcoming CMB experiments.} For lower noise levels of 2 and 1 $\mu$K-arcmin—which are representative of the capabilities of LiteBIRD and CMB-S4—the sensitivity improvements after delensing range from 25\% to 40\%. These results underscore the potential of delensing to enhance the detectability of subtle parity-violating signals in the CMB polarization.

In the anisotropic case, we reconstructed the birefringence angle using a quadratic estimator applied to the CMB polarization fields over the multipole range $2 \leq \ell \leq 1024$, for a simulation with a noise level of 1 $\mu$K-arcmin. Recognizing that gravitational lensing introduces additional biases (including contributions to the {$N^\alpha_{(1)}$} term), we compared our results with lensed Gaussian simulations to compute a bias spectrum that captures these higher-order effects. Although we estimated and mitigated the lensing bias, our likelihood analysis was restricted to multipoles below $\ell = 100$ to minimize the impact of the unmitigated {$N^\alpha_{(1)}$} bias. For the anisotropic case, delensing reduces the {$N^\alpha_{(0)}$} bias by about 50\%, and the posterior distribution of the amplitude $A_{\rm CB}$ improves by approximately 30\% following delensing.

In both scenarios, upcoming experiments—such as the Large Aperture Telescope of the Simons Observatory and CMB-S4—are expected to probe smaller angular scales, where delensing can further enhance the statistical significance of CB measurements. Future work will aim to incorporate additional complexities, including foreground contamination, instrumental calibration uncertainties, anisotropic noise, and partial sky observations, as well as to address residual biases by incorporating realization-dependent {$N^\alpha_{(0)}$ and $N^\alpha_{(1)}$} terms. Moreover, we plan to implement fully iterative estimators to further refine our results. Overall, our study demonstrates that delensing is a powerful tool for enhancing the sensitivity of CB measurements, paving the way for robust tests of new physics beyond the Standard Model in forthcoming CMB observations.  The analysis pipeline is publicly available at \url{https://github.com/antolonappan/DelensCoBi}.

\section*{Acknowledgments}
{We express deep gratitude to the anonymous referee for their insightful and constructive feedback, which greatly enhanced the clarity and quality of this manuscript. We are also thankful to Toshiya Namikawa, Sebastian Belkner, Kevin Crowley, Kam Arnold, and Brian Keating for their valuable discussions and  feedback. This research was conducted using the Popeye cluster at the Flatiron Institute, hosted by the San Diego Supercomputer Center. We acknowledge the use of \texttt{NumPy}, \texttt{SciPy}, \texttt{Matplotlib}, \texttt{Healpy}, \texttt{CAMB}, \texttt{emcee}, \texttt{Lenspyx}, \texttt{Plancklens} and \texttt{cmblensplus}, a community-developed core Python packages, which were instrumental in this work.}

\bibliographystyle{jcap/JHEP}
\bibliography{reference}
\end{document}